# How far does scientific community look back?

Xianwen Wang[1,2,3]*, Zhi Wang[1,2], Wenli Mao[1,2], Chen Liu[1,2]

[1] WISE Lab, Faculty of Humanities and Social Sciences, Dalian University of Technology, Dalian 116085, China.

[2] School of Public Administration and Law, Dalian University of Technology, Dalian 116085, China.

[3] DUT- Drexel Joint Institute for the Study of Knowledge Visualization and Scientific Discovery, Dalian University of Technology, Dalian 116085, China.

\* Corresponding author.
Email address: xianwenwang@dlut.edu.cn

**Abstract**: How does the published scientific literature used by scientific community? Many previous studies make analysis on the static usage data. In this research, we propose the concept of dynamic usage data. Based on the platform of realtime.springer.com, we have been monitoring and recording the dynamic usage data of *Scientometrics* articles round the clock. Our analysis find that papers published in recent four years have many more downloads than papers published four years ago. According to our quantitative calculation, papers down-loaded on one day have an average lifetime of 4.1 years approximately. Classic papers are still being downloaded frequently even long after their publication. Additionally, we find that social media may reboot the attention of old scientific literature in a short time.

**Keywords**：article usage; altmertics; download; static usage data; dynamic usage data; lifetime of scientific literature

## 1. Introduction

The visibility of paper has long been reflected mostly in the reference of citing papers. A lot of scientometrics indicators and methods are introduced based on the references, i.e. citation analysis. Price is the first people to study the age of references. In his classic article, the Price index is defined as the "the proportion of the references that are to the last five years of literature" (Price, 1970). Price index has been improved by many later studies (Egghe, 1997, 2010; Egghe & RavichandraRao, 2002; Glänzel & Schoepflin, 1995). The aging process of scientific literature is described with a mathematical model and studied based on large-scale data (van Raan, 2000). The temporal evolution of the aging phenomenon over more than 100 years of scientific activity is studied, when the results show that the age of cited references has risen continuously since the mid-1960s (Larivière & Archambault, 2008). Other indicators rooted in references include cited/citing half-life, impact factor, immediate index, etc. The aging degree in different scientific fields maybe different, e.g., previous studies find that chemists rely heavily on current journals (Hurd, Weller, & Curtis, 1992), when mathematicians make more use of older references (Garfield,

1983).

Besides the citations reflected in references, there are also many other ways to embody the value of scientific literature, i.e. the full-text downloads, the altmetrics indicators (Bar-Ilan et al., 2012; Priem, Taraborelli, Groth, & Neylon, 2010). Compared with other classical impact indicator, e.g. citations, and emerging impact indicator, e.g. altmetrics, the download times of paper, as the most front-end impact indicator, may provide new insights on the usage of scientific works and their impact on scientific community.

Bibliometrics is defined as the quantitative study of publication data on the basis of citation and text analysis (Broadus,1987). Now the definition is expanded because of the growing role of usage data, bibliometrics can include studies based on usage data (Kurtz & Bollen, 2010).

The ease of finding and accessing scientific scholarly articles on the web make that most academic researchers primarily use electronic access for searching, retrieving and reading content, which may conduct significant change to the information seeking behavior of science researchers (Hemminger, Lu, Vaughan, & Adams, 2007).

In the Pre-Electronic Era, the article is attached with a specific issue and could not be separated, which means that people isn't able to borrow an article, instead of an issue, from the library. So, it's impossible to count the usage of one specific article, when digital library makes it possible. However, in earlier periods, it is difficult to know how many times a paper has been downloaded. Although publishers have these usage log data in their servers, they may not make it public. Nowadays, more and more scientific publishers have the willingness to provide article-level metrics data to public. Researchers have made good tries to explore the relationship between the article usage and citation, strong correlation between the number of downloads and citation frequencies is identified and download counts can be used as early performance indicators of subsequent academic impact for papers (Brody, Harnad, & Carr, 2006; Harnad & Brody, 2004; Perneger, 2004; Schloegl & Gorraiz, 2010).

However, in these previous studies, only the total downloaded time, which could be called static usage data, is used to make the analysis. During the recent years, more and more publishers began to provide detailed article-level usage data, from which we know not only how many times one paper has been downloaded, but also what time the downloading occurs. We may call it a kind of dynamic data. For example, *PLOS*, *IEEE* and *PNAS* provide the total article views and cumulative views in each month for each paper. In October, 2012, *Nature* began to provide daily page views counts for each research paper(Wang, Mao, Xu, & Zhang, 2014). In December, 2010, Springer Verlag launched an analytic platform, realtime.springer.com. It aggregates the data on downloads of Springer journal articles, images, book chapters, and protocols in real time from scientists all over the world, and displays the downloading in a variety of visualizations. For each journal and book, the platform also displays the latest downloaded items realtimely, including the title and author of the downloaded item, etc.(http://realtime.springer.com/feed).

Using this kind of dynamic usage data, it is possible for us to make more detailed analysis on how a scientific paper been used after its publishing. Since March 1, 2012, we have been monitoring the realtime downloads and collecting dynamic usage data from realtime.springer.com round the clock for more than one year so far. Using the data collected from real-time.springer.com, we have conducted several studies, i.e., exploring scientists' working timetable (Wang et al., 2012; Wang, Peng, et al., 2013; Wang, Wang, & Xu, 2013), tracing the real-time research trend (Wang, Wang, et al., 2013). In this research, taking *Scientometrics* for example, we focus on the dynamic usage data of every single *Scientometrics* paper. Our research question is, does the usage of scientific scholarly articles has any identifiable patterns or rules?

## 2. Data

### 2.1 Dynamic usage data

From the platform of realtime.springer.com, we can know that what paper is being downloaded right now. For *Scientometrics* journal, the realtime downloads data could be collected and extracted from http://realtime.springer.com/feed. The data is monitored and recorded round the clock day by day. Every time when a download occurs, the downloaded time(Greenwich Time), the title of article, the authors and the DOI of the article are recorded immediately. A local database is designed in SQL Server 2008 to store, process and analyze the collected raw data.

### 2.2 Publish data

From 1978 to July 1, 2013, 96 Volumes, 295 Issues and 3739 Articles have been published in *Scientometrics*. We need to confirm the publish date for each article. Before May, 2007, all papers in one issue have the same publish date, which are just the issue date. However, articles published in and after May, 2007 may have two publish dates, which are publish online date and cover date. So, for the papers published in Volume 71, Issue 1 and before, the cover date of the issue is considered as the publish date of all articles in the whole issue. For papers published in Volume 71 Issue 2 and after, we have to confirm the Publish Online date for each article one by one.

## 3. Results

### 3.1 Usage statistics

2942 *Scientometrics* articles have been downloaded 40,296 times during the time period from April 11, 2013 to June 30,2013. As of the end of June, 2013, there are 3748 articles published in *Scientometrics*, including online first articles (articles not assigned to an issue) and published issue articles. So, 2942 downloaded articles accounts for 78.50% of the total 3748 *Scientoemtrics* articles.

The investigation time period is divided into several 7-day timespans, e.g., April 11–April 17. However, from May 9 to May 15, most of the data is missing because the platform of realtime.springer.com has some problems. So the data from May9 to May 15 is not included in this study. As Table 1 shows, the numbers of downloaded articles vary from 1307 to 1628,which accounts 35.27–43.63% of all published *Scientometrics* articles.

Table 1 Number of downloaded articles in one week

|  | Number of downloaded articles | Number of published articles | Percent |
| --- | --- | --- | --- |
| April 11- April 17 | 1385 | 3699 | 37.44% |
| April 18-April 24 | 1407 | 3699 | 38.04% |
| April 25-May 1 | 1424 | 3706 | 38.42% |
| May 2-May 8 | 1307 | 3706 | 35.27% |
| May 16-May 22 | 1336 | 3728 | 35.84% |
| May 23-May 29 | 1466 | 3728 | 39.32% |
| May 30-June 5 | 1628 | 3731 | 43.63% |
| June 6-June 12 | 1517 | 3738 | 40.58% |
| June 13-June 19 | 1512 | 3744 | 40.38% |
| June 20-June 26 | 1457 | 3745 | 38.91% |

### 3.2 Publish-download time interval

For the downloaded articles, we need to check the publish date for each item. As Fig. 1 shows, the green fine lines indicate downloads on specific days during the investigation time period, when the thick red line shows the average downloads of all days. The publish dates mentioned here are confirmed by the cover date of each issue. Most curves has the similar shape.

For those articles published in 2009, they have 17.08 average downloads each day. However, downloads of articles in2010 increase greatly to 81.13 times per day. For the articles published in 2011, the average downloading number has a small drop to 68.36 times per day. There are only 173 articles published from January to June, 2013, however, these 173 articles get the most total downloads compared with articles published in other years. Articles of 2013 has 110.73 average daily downloads, when 270 articles published in 2012 get the second most downloads, 85.80 times each day.

Taking the year 2009 as the dividing point, articles published in and before 2009 get many less downloads than articles published after 2009, which means that articles published in recent four years get much more attention than older articles. The distribution of downloads according to the publish years could be seen clearly from Fig. 1.

For *Scientometrics* articles published after 2007, there are usually two publish dates, the Publish Online date and issued date. Here we consider the online date for articles after 2007 as the publish date, and for articles published in and before 2007, we still use the cover date as the publish date. Now the download curve in Fig. 1 could be redrawn in Fig. 2. The blue fine lines indicate the downloads on specific days during the investigation time period, when the thick red line shows the average downloads of all days.

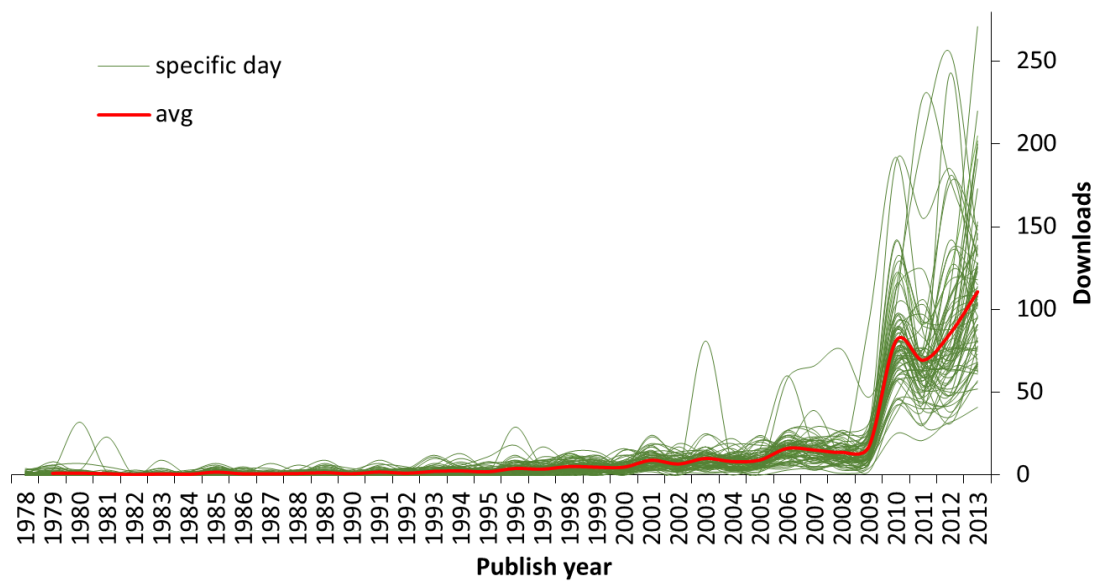

**Fig. 1** Downloads of articles published in past years according to the cover date

Here we still focus on the daily average downloads, compared with Fig. 1, the shape of the right part of the curve in Fig. 2 is very different. 135 articles published in 2013 only get 43.15 downloads per day. 335 articles published in 2012 have the most downloads, 160.56 per day. The year 2009 could be still regarded as the demarcation point. Articles published in recent four years still get many more downloads than articles published four years ago. Different from the results shown in Fig. 1, articles in 2012 have the most downloads. However, the two results in Figs. 1 and 2 are not conflict. Because of the publish delay from the Publish Online date to issued data (usually several months), many articles published online in 2012are issued in 2013.

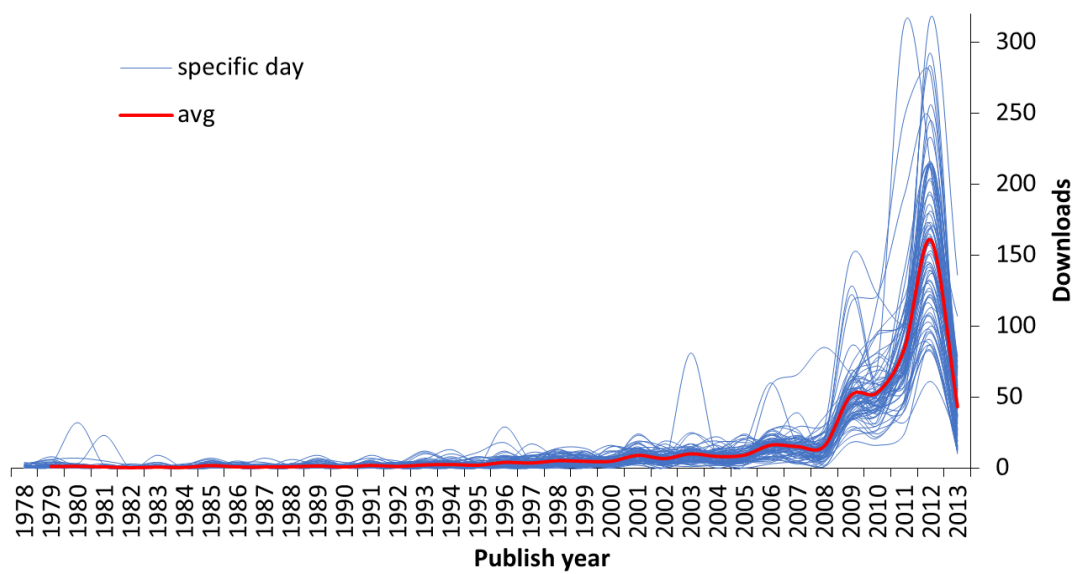

**Fig. 2** Downloads of articles published in past years according to the publish online date

It seems that new issued articles get more downloads than new published online articles. For the latter ones, it may get some attention by those researchers who visit the journal website regularly or use RSS and other feed tools to read the newest articles. However, for each new issue, Springer would send emails to alerts subscribers about the journal tables of contents of new journal issue, namely Springer Alerts, as Fig. 3 shows. We think this policy is the most likely reason to promote the downloads of new issued articles.

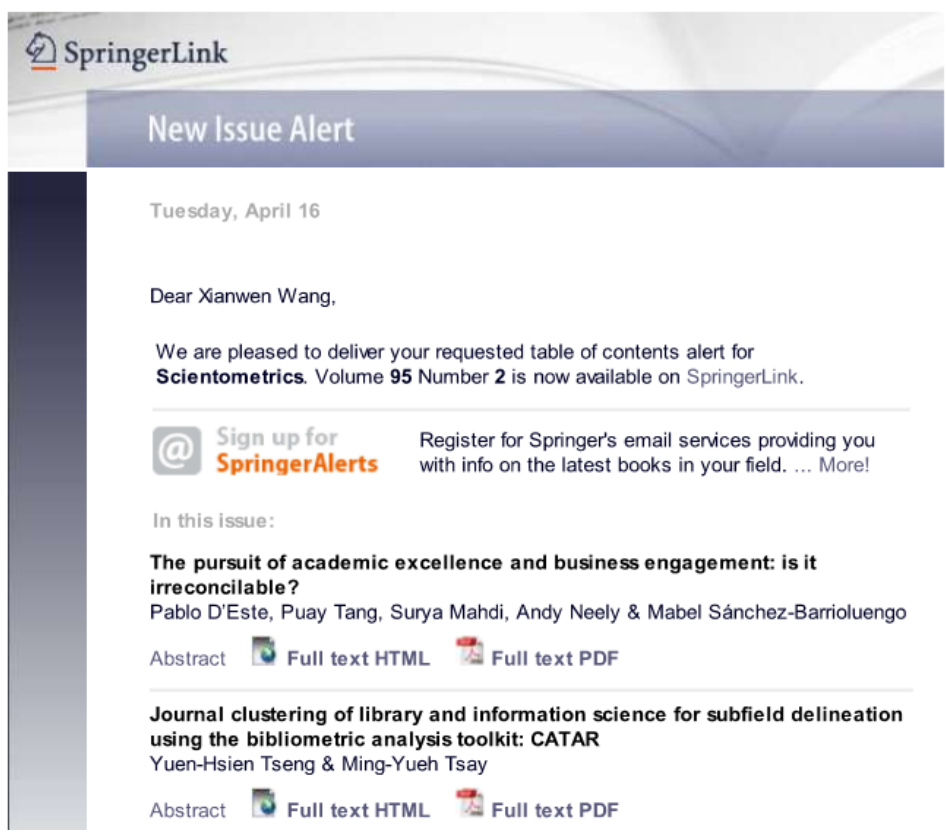

Fig. 3 New issue alert email of *Scientometrics*

### 3.3 Usage statistics of a new issue

Three new issues are published during our study period. The data from May 9–15 is missing because the platform ofrealtime.springer.com does not work well, when Mid-May is the publish date of Volume 95 Issue 3. And Volume 96 Issue 1 is published on June 17, the study period is not long enough to make a complete assessment. So, here we only choose Volume95 Issue 2 of *Scientometrics* as a specific example to analyze the downloading trend when a new issue is published, as Fig. 4 shows.

The new issue alert email of Volume 95 Issue 2 is sent on April 17, 2013. However, some downloads occur before April17. On April 16, the number of downloads is 45.

There is a considerable increase on April 17, when the new issue articles have been downloaded 138 times. After that the curve falls rapidly. On April 18, the number is only 58.

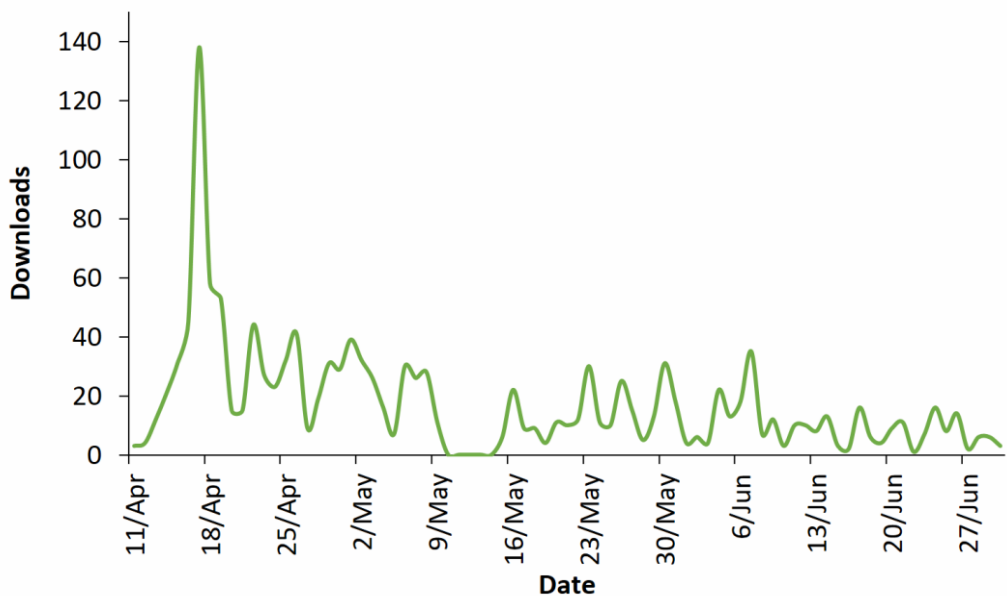

**Fig. 4** Download statistics of Volume 95 Issue 2 of *Scientometrics*

### 3.4 The lifetime of downloaded literature

The lifetime (publish–download time interval) of each downloaded paper on each day is calculated to measure the lifetime of downloaded literature.

$$L_d = \frac{\sum_{i=1}^{n}(L_i \times t_i)}{\sum_{i=1}^{n} t_i}$$

$L_d$ indicates the average lifetime on one day, $L_i$ is the lifetime of the downloaded paper $i$, when $t_i$ implies the total downloaded times of paper $i$ on the same day, and $n$ is the number of downloaded papers.

In Table 2, we show an example of how to calculate the average lifetime on a specific day. On April 12, 2013, 432 *Scientometrics* papers have been downloaded 656 times. There are two kinds of lifetime, one based on issue date and the other is online date, we checked the both data and choose the greater value. The last column is the result of the lifetime multiply the downloaded times.

The total lifetime on April 12 is summarized by the values in the last column, and the result is 796,509. So, divide 796,509 by 656, the result 1214.19 is the average lifetime on April 12. For those articles not assigned to an issue, the lifetime (issue date) is NULL, when the lifetime (online date) is NULL for articles published before Volume 71 Issue 2.

**Table 2** Data on April 12, 2013

| Rank (n) | DOI | Times ($t_i$) | Lifetime (issue date) | Lifetime (online date) | Lifetime ($L_i$) | $L_i \times t_i$ |
|---|---|---|---|---|---|---|
| 1 | 10.1007/s11192-013-1013-9 | 14 | NULL | 0 | 0 | 0 |

| 2 | 10.1007/s11192-012-0829-z | 12 | 32 | 237 | 237 | 2844 |
| 3 | 10.1007/s11192-012-0861-z | 10 | 32 | 185 | 185 | 1850 |
| 4 | 10.1007/s11192-011-0420-z | 8 | 608 | 670 | 670 | 5360 |
| 5 | 10.1007/s11192-005-0255-6 | 8 | 2811 | NULL | 2811 | 22,488 |
| 6 | 10.1007/s11192-012-0854-y | 6 | 32 | 199 | 199 | 1194 |
| … | … | … | … | … | … | |
| 432 | | | 656 | | | 796,509 |

We calculate the daily average lifetime of downloaded papers for 75 days. The mean lifetime is 1533.73 days, when the median number is 1555.25. In Fig. 1, we find that papers published after 2009 get many more downloads than papers published in and before 2009. So, the average lifetime calculated here is quite consistent with the results in Fig. 1.

Fig. 5 shows the distribution of daily average lifetime on the 74 days. Most data belongs to the range of 1300–1650.

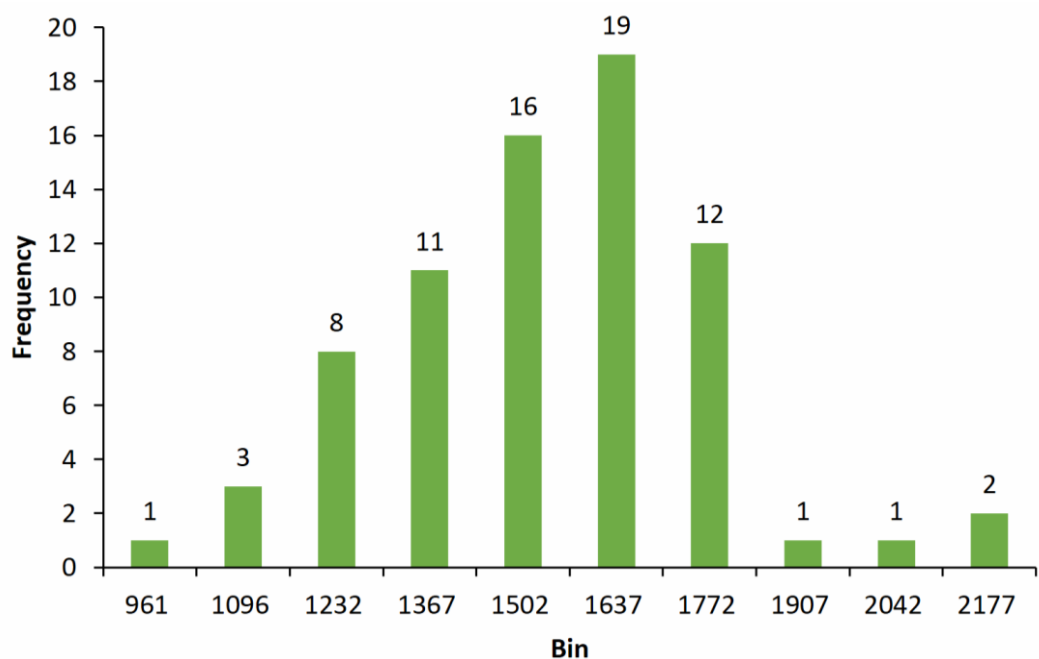

**Fig. 5** Histogram of daily average lifetime of downloaded literature

### 3.5 Highly downloaded papers

As mentioned above, papers published in recent four years tend to get the most downloads. Fig. 6 shows the downloads and citations of papers published in the year of 2005. The total downloaded times in the study period correlates well with the cited times, when the correlation coefficient is 0.85.

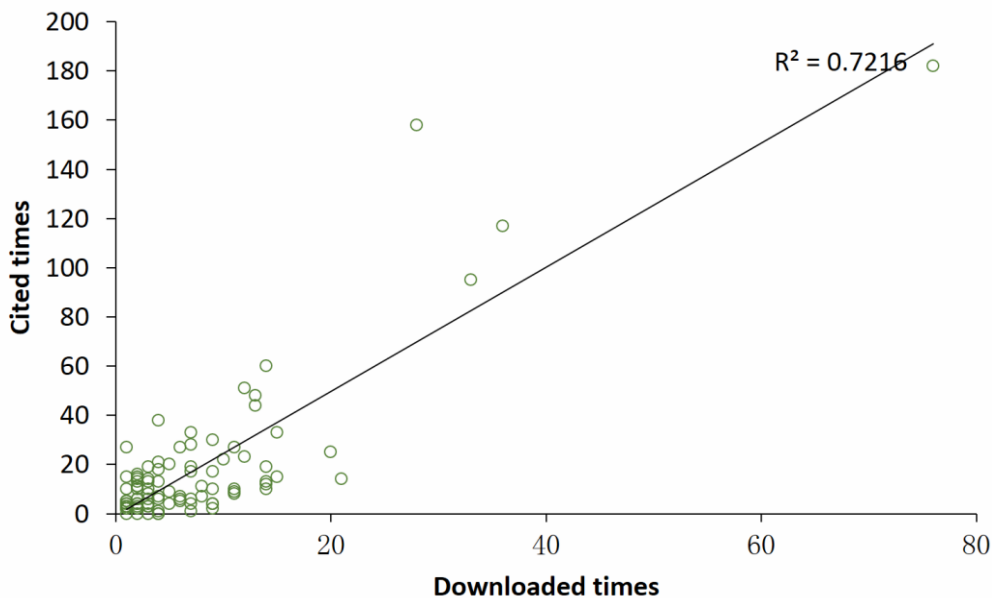

**Fig. 6** Scatter plot of downloads and citations of papers published in 2005

The conclusion in Fig. 6 could be applied to papers published in other years. However, we also find a special case with high download times and not so much citations. The paper "Letter to the Editor: A global comment on scientific publications, productivity, people, and beer" was published July 18, 2009. It has been cited 1 time in Web of Science and downloaded more than 320 times during the investigation time period. The daily downloaded times of this paper are shown in Fig. 7. It is very surprising because the paper has so much attention after four years of its publication. Further investigation reveals that the social media has played the major role in promoting the download greatly in 2013. In April 2013, the main conclusion of this paper is posted to the web, including Figshare (April 2), arstechnica.com (April 14), imperial.tab.co.uk (April 15), and also twitter, Facebook, Google plus, etc. The wide spread of the main standpoints and conclusions on the social web cause the highly download of this paper.

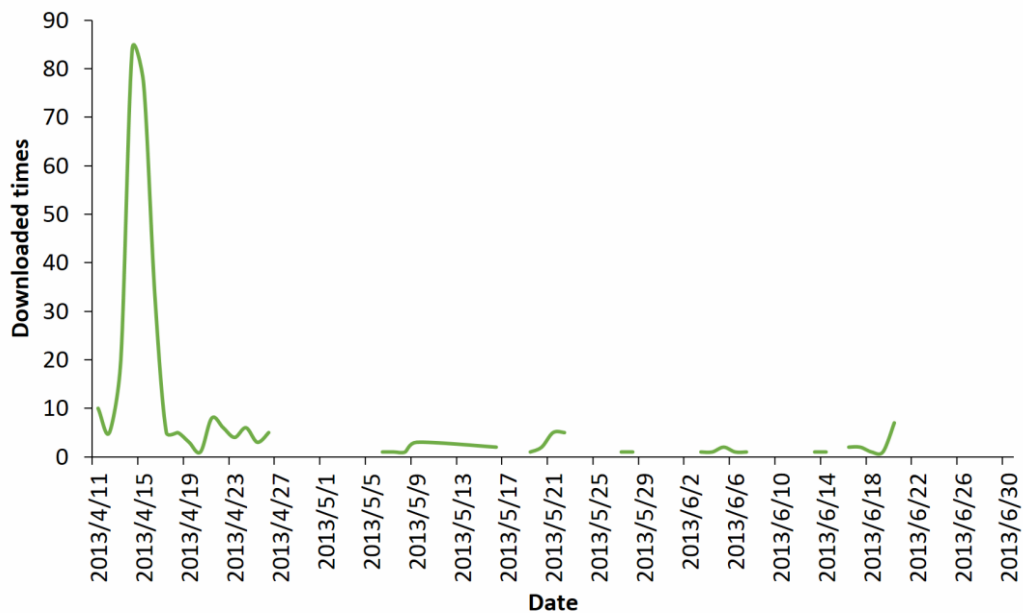

**Fig. 7** Downloaded times of paper 10.1007/s11192-009-0077-z

Actually, this is the second case we find about the old literature regains people's attention. According to a recent summary, top ten articles are selected which have received the most online attention in the whole year of 2012 based on the altmetrics data (Liu, 2012). Nine of the ten highly focused papers are published in 2012 or 2011. However, one paper published in 1996 has got over 1600 tweets in August, 2012. Why a paper published in 1996 gets so much attention after 16 years? The cause is the inappropriate remarks about abortion by Todd Akin, a Republican nominee for the U.S. Senate. He shocked people when he said in a television interview in August, 2012, "If it's a legitimate rape, the female body has ways to try to shut that whole thing down." The comments triggered off immediate public backlash (Liu, 2012). Then the 1996 study published in the *American Journal of Obstetrics and Gynecology* was found and disseminated as convincing refutation of the improper remarks. The main finding in this 1996 paper is that there are 23, 101 pregnancies result from rape each year in America(Holmes, Resnick, Kilpatrick, & Best, 1996).

## 4. Discussion

In this study, we analyze the dynamic usage of papers published in *Scientometrics*. We find that papers published in the recent four years have got many more downloads than older ones. According to our quantitative calculation of publish–download time interval of downloaded items, the daily average lifetime of downloaded papers is about 1500 days, which equal to approximate 4.1 years. When a new issue published, it will get much attention in one or two days. We think this phenomena is because of the email advertising policy by Springer, namely Springer Alerts. Highly cited papers are still being downloaded a lot even many years after their publication. Social media may reboot the attention of old scientific literature in a short time, even for those not so highly cited papers.

The limitations in this research are that the investigation time period is only two and half months, and only one journal is studied. We are still collecting the data round the clock, and will extent our future studies in this direction with data of a longer investigation time period, i.e., two years or even longer. And, different topics in scientometrics and different scientific fields would be analyzed as well.

## Acknowledgements

We would like to thank Chaomei Chen and Zeyuan Liu for providing us useful suggestions. The work was supported by the project of "National Natural Science Foundation of China" (61301227) and the project of "Social Science Foundation of China" (10CZX011) .